\def\ANON{0} 
\def\USELINKS{1} 
\def\AAM{0} 
\def\PAGENUMS{0} 
\def\LINKBOXES{1} 
\def\ARXIV{1} 
\def\CR{0} 

\ifnum\CR=1
    \def\LINKBOXES{0} 
    \def\PAGENUMS{0} 
    \def\AAM{0} 
    \def\ANON{0} 
    \def\USELINKS{1} 
\else
    \ifnum\ARXIV=1
        \def\AAM{1} 
        \def\LINKBOXES{0} 
        \def\PAGENUMS{1} 
        \def\USELINKS{1} 
        \def\ANON{0} 
    \fi
\fi

\documentclass[conference]{IEEEtran}
\ifnum\USELINKS=1
    \usepackage[hyphens]{url} 

    \ifnum\LINKBOXES=1
        \usepackage[breaklinks]{hyperref}
    \else
        \usepackage[breaklinks,hidelinks]{hyperref}
    \fi
\else
    \usepackage[hyphens]{url} 
\fi

\usepackage{amsmath}
\usepackage{graphicx}
\usepackage{xcolor}
\usepackage{longtable}
\usepackage{diagbox}
\usepackage{multirow}
\usepackage{booktabs} 
\usepackage{flushend}
\usepackage{threeparttable}
\usepackage{tabularx}
\usepackage{siunitx}
\usepackage{fancyhdr} 
\usepackage{lastpage} 

\ifCLASSOPTIONcompsoc
  \usepackage[nocompress]{cite}
\else
  \usepackage{cite}
\fi

\everypar{\looseness=-1}

\begin{document}
%
\title{zkMixer: A Configurable Zero-Knowledge Mixer with Anti-Money Laundering Consensus Protocols}

\ifnum\ANON=1
    \author{\IEEEauthorblockN{Anon}
    \IEEEauthorblockA{\\ \\ \\ \\}
    }
\else

    
    \author{\IEEEauthorblockN{Theodoros Constantinides}
    \IEEEauthorblockA{School of Computer Science\\
    University of Bristol, Bristol, UK\\
    Email: theodoros.constantinides@bristol.ac.uk}
    \and
    \IEEEauthorblockN{John Cartlidge}
    \IEEEauthorblockA{School of Engineering Mathematics and Technology\\
    University of Bristol, Bristol, UK\\
    Email: john.cartlidge@bristol.ac.uk}}
\fi


\maketitle

\ifnum\PAGENUMS=1
    \thispagestyle{fancy}
    \pagestyle{fancy}
    \fancyfoot[C]{\fontsize{9}{10} \selectfont Page \thepage ~of {\hypersetup{hidelinks}\pageref{LastPage}}}
    \ifnum\AAM=1 
        \fancyhead[C]{\fontsize{9}{10} \selectfont Accepted author manuscript: {\em 7th IEEE International Conference on Decentralized Applications and Infrastructures (IEEE DAPPS 2025)}} 
    \fi
\fi

\begin{abstract}
We introduce a zero-knowledge cryptocurrency mixer framework that allows groups of users to set up a mixing pool with configurable governance conditions, configurable deposit delays, and the ability to refund or confiscate deposits if it is suspected that funds originate from crime. Using a consensus process, group participants can monitor inputs to the mixer and determine whether the inputs satisfy the mixer conditions. If a deposit is accepted by the group, it will enter the mixer and become untraceable. If it is not accepted, the verifiers can freeze the deposit and collectively vote to either refund the deposit back to the user, or confiscate the deposit and send it to a different user. This behaviour can be used to examine deposits, determine if they originate from a legitimate source, and if not, return deposits to victims of crime. 
\end{abstract}


%
\IEEEpeerreviewmaketitle

\section{Introduction}
\noindent
Most cryptocurrencies, such as Bitcoin, were created to facilitate the exchange of value, i.e., act as money. However, the transparent and immutable nature of blockchains means that most blockchain systems only provide pseudo-anonymity, which is less anonymity than traditional cash.  

There are many ways for users to increase their privacy when transacting using cryptocurrencies, each with varying beneficial and detrimental properties. Common approaches include using privacy-focused cryptocurrencies, such as Monero \cite{monero} and Zcash \cite{zerocash}; moving funds through different wallets (wallet hopping) and decentralized finance (DeFi) applications to complicate transaction tracing; moving funds between different blockchains using bridges; using centralized services such as over-the-counter (OTC) brokers or exchanges; and interacting with mixing services.

Mixing services -- also known as mixers, tumblers, or blenders -- are common in the blockchain space and are designed to increase privacy by breaking the linkability of transactions. As the name suggests, mixers offer anonymity by mixing the funds of many users. This grants anonymity to a user by adding them to the anonymity set of all the users of the mixing service. Unfortunately, the anonymity provided by mixers means that they are commonly used by criminals looking to launder illicit funds. Since 2017, North Korean hackers have stolen more than \$6 billion of crypto assets, with much of this laundered using mixers and other layering techniques  \cite{bydit}.\footnote{While we were writing this paper (21 Feb 2025), North Korean hackers conducted the largest known theft of any kind, stealing \$1.46 billion from Bybit, a Dubai-based centralized cryptocurrency exchange \cite{bydit}.} 

In this paper, we tackle the problem of money laundering through mixers by introducing a novel mixer protocol that uses zero-knowledge proofs (ZKP) and a consensus process to monitor mixer inputs and decide whether to allow, refuse, or confiscate deposits. Our protocol enables users to establish complex criteria on who is allowed to deposit funds to the mixer, while preserving user/deposit anonymity. 

\section{Background}
\subsection{Illicit Activity on the Blockchain}
\label{sec:crime}
\noindent
The blockchain analysis company, Chainalysis, has concretely identified \$40.9 billion of illicit on-chain activity in 2024, and estimates that the true amount could be closer to \$51 billion \cite{2025report}, which corresponds to 0.14\% of the total annual transaction volume.  Most of the illicit activity identified (63\%) involved the use of stablecoins, and included transfers to sanctioned entities/jurisdictions (\$15.8B), cybercrime (\$10.8B), frauds and scams (\$9.9B), market manipulation (\$2.6B), stolen funds (\$2.2B), darknet markets (\$2B), and ransomware (\$0.8B) \cite{2025report}.

Mixers are not immune from crime, and their privacy-enhancing properties can attract bad actors who are looking to launder stolen funds or proceeds of crime. Chainalysis tracked \$31.5B (2022) and \$22.2B (2023) of laundered cryptocurrencies \cite{2024report}. Of these, \$1.01B (2022) and \$0.5B (2023) were moved through mixers. To put this into perspective, in 2022 the total amount moved through mixers was \$7.8B \cite{mixerUsage}, and Tornado Cash, which used to be one of the most popular mixers, had more than 25\% of its deposits linked to illicit activities \cite{2025report}.

Criminals using mixers also degrade the experience of mixers for legitimate users. Legitimate users, who only wish to use mixers to gain anonymity, must accept the stigma of holding funds that are potentially associated with criminal activities. Even worse, some individuals, businesses, or even smart contracts may choose not to accept funds originating from certain mixers \cite{dusting}. This limits the use of mixer-anonymized cryptocurrencies and, in some sense, funds that are passed through a mixer become tainted.

To combat crime, the governments of the US and several European countries have started to take down servers that host mixers, arrest their operators, and even sanction the mixers themselves. Notable server seizures include Bestmixer.io \cite{bestmixer}, ChipMixer \cite{chip}, and Sinbad.io \cite{sinbad}. Notable arrests include the operators of Bitcoin Fog \cite{fog}, Helix \cite{helix}, Samourai Wallet \cite{samourai}, and the creators of Tornado Cash \cite{tornadoArrest1}. The arrest of the founders of Tornado Cash is interesting because, unlike the other examples presented here, Tornado Cash is a decentralized smart contract that the founders do not directly control. The Office of Foreign Assets Control (OFAC) of the US Treasury has also sanctioned several mixers, including Blender.io \cite{blender}, Tornado Cash \cite{tornadoSanction1}, and Sinbad.io \cite{sinbadSanction}. Again, the case of Tornado Cash is interesting, as sanctions were later overturned by the US Court of Appeals, stating, ``we hold that Tornado Cash’s immutable smart contracts (the lines of privacy-enabling software code) are not the `property' of a foreign national or entity, meaning (1) they cannot be blocked under IEEPA, and (2) OFAC overstepped its congressionally defined authority'' \cite{tornadoAppeal}. Funds originating from a sanctioned mixer can also be used for a malicious purpose, such that individuals are sent small amounts of these funds -- a process known as ``dusting'' -- in order to put the receivers of funds in breach of sanctions.

\subsection{Types of Mixers}
\begin{figure}[tb]
    \centering
    \includegraphics[width=\linewidth]{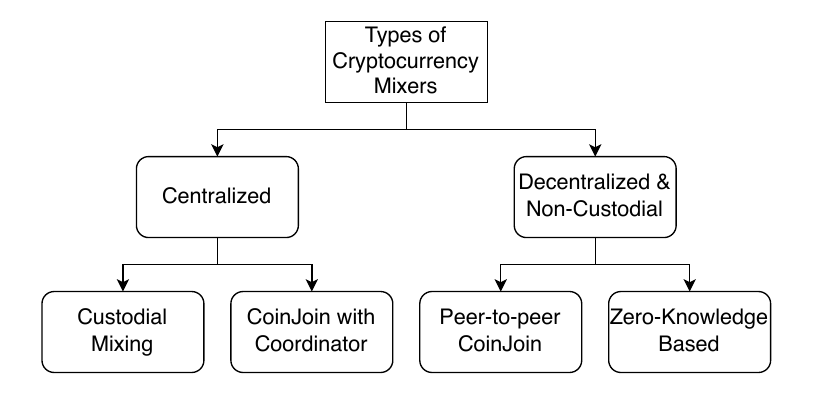}
    \caption{Categorization of common cryptocurrency mixers.}
    \label{fig:mixer-types}
\end{figure}

\noindent
Fig.~\ref{fig:mixer-types} presents a classification of the most common mixer designs. Centralized mixers rely on an administrator/coordinator to perform the mixing. This architecture has many disadvantages: the operator has full knowledge of inputs and outputs, which can later be used to break anonymity; and in the case of custodial mixing services, the operator has custody over funds. For these reasons, more advanced decentralized mixers were developed.

Decentralized mixers do not depend on an operator but on cryptography. Specifically, most of these mixers either use transactions with multiple inputs and outputs --- an idea first introduced in early 2013 by Maxwell and known as CoinJoins \cite{coinjoin1, coinjoin2} --- or Zero-knowledge proofs (ZKPs) \cite{ZKP}. A CoinJoin on its own requires a coordinator, but this can be avoided by using a peer-to-peer network. However, in both cases, the coordinator, or the rest of the peers, have complete knowledge of all transaction routing. This can be prevented by using blind signatures \cite{blind} or more complex protocols, such as CoinShuffle \cite{coinShuffle}. CoinJoins are simpler than ZKPs and are mainly used in blockchains such as Bitcoin, which do not support complex operations. However, the anonymity set is usually limited. Zero-knowledge (ZK) based mixers provide better anonymity, as their anonymity sets include all users of the mixer, and not just those of a single CoinJoin instance. However, ZKPs require more complex operations and are therefore only suitable for smart contract-enabled blockchains such as Ethereum. 

ZK-based mixers allow users to deposit funds, along with some secret, onto a smart contract and then later withdraw those funds by proving that they know the secret. Importantly, during withdrawal, users do not reveal which secret they know; using a ZKP, they only reveal that they know one of the secrets of all the mixer users. This ensures that users get the full benefit of the anonymity set of the mixer and deposits cannot be stolen by others, as long as the code works as intended and the secret is not shared. 

\subsection{Zero-Knowledge Proofs}
\noindent
A zero-knowledge proof (ZKP) allows a party (the prover) to prove a statement to another party (the verifier) without revealing anything apart from the statement that is being proven. Although early ZKPs were impractical for use in smart contracts, recent improvements, such as the introduction of zero-knowledge succinct non-interactive arguments of knowledge (zk-SNARKs) \cite{zkSnark}, have enabled practical blockchain applications (see Nitulescu \cite{zkSnarkSurvey} for a history of ZKP systems). Today, zk-SNARK variations are commonly used in decentralized blockchain-based systems, e.g., Pinocchio \cite{Pinocchio}, Groth16 \cite{groth16}, and PlonK \cite{PLONK} (see Liang et al. \cite{zkSnarkSurvey2} for a survey of common systems and tools for zk-SNARKs).

\subsection{Zero-Knowledge Cryptocurrency Mixers}\label{sec:ZKMixer}
\begin{figure}[tb]
    \centering
    \includegraphics[width=0.95\linewidth]{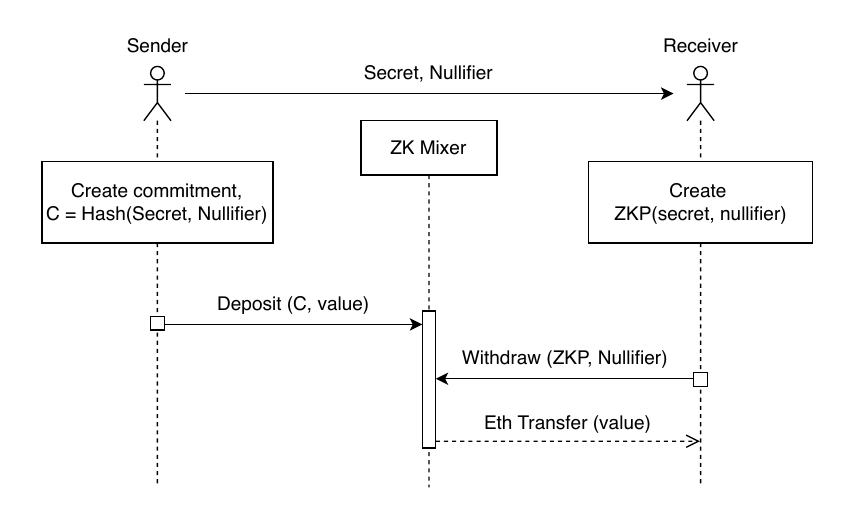}
    \caption{Example of simplified operation of a ZK mixer. The sender is able to transact with the receiver indirectly through the mixer without the possibility of linking the two. Note that it is possible for the sender and receiver to be two different addresses controlled by the same user.}
    \label{fig:mixer}
\end{figure}

\noindent
ZK cryptocurrency mixers work by breaking the link between the sender and the recipient of a transaction. The main idea is that, instead of sending funds directly to a recipient, the sender first sends the funds to a smart contract and allows the recipient (which could be a fresh address controlled by the sender) to later withdraw them. Of course, implementing this naively, e.g., by using a mapping holding user balances, would not break the link between the sender and the recipient; it would at best make observing this connection slightly more difficult. This is where ZKPs enter the picture. Using ZKPs, a system can be designed so that there is no way to link the input with the output of a transaction passing through the mixer smart contract. To explain how this works, we will describe a simplified version of Tornado Cash \cite{tornado}, as most ZK cryptocurrency mixers follow a similar design (see Fig.~\ref{fig:mixer}). 

ZK mixer protocols have two phases, the deposit and the withdrawal. Before the deposit, the sender creates a hash of two large random values, the secret and the nullifier. This hash is known as a commitment. Then, during the deposit, the user sends a fixed amount of funds to the mixer along with their commitment. The smart contract of the mixer then stores the commitment in a Merkle tree \cite{MerkleTree} and keeps the funds until a withdrawal is made. Before the withdrawal can take place, the recipient must create a ZKP in the form of a zk-SNARK, to prove that the recipient knows the secret and nullifier that were used for one of the commitments that are stored in the mixer's Merkle tree. Importantly, this zk-SNARK does not reveal which commitment the recipient knows, as this would be enough to link the recipient to the sender; it only reveals that the recipient knows one of them. The recipient then sends this zk-SNARK along with the value of the nullifier to the mixer. The mixer then verifies that the zk-SNARK is correct and checks that the provided nullifier has not been used before. If both of these conditions are satisfied, the mixer stores the provided nullifier and sends the funds to the recipient. 

In the previous description, we glossed over some important details. It is worthwhile explaining these details, as similar design choices will be used later in our own protocol. First, the commitments use both a secret and a nullifier. This is because we need to keep the content of the commitment secret to avoid linking of the sender and receiver, but we also need to make sure that commitments cannot be spent more than once. The nullifier is later used to identify if a commitment was spent. Without the nullifier check, a user could deposit once and then withdraw multiple times from the mixer. By having two parts in a commitment and only revealing one, a user can be sure that their identity will not be revealed, while the mixer can be sure that the recipient is not trying to spend the same commitment twice. Second, we mentioned that users need to deposit fixed amounts of funds. The reason for this requirement is that if the same mixer allowed for different amounts, then it would be easy to link the sender and the recipient. For example, if there is a unique deposit of 0.123 Eth in the mixer, it will be obvious that the recipient is linked to the person who deposited that specific amount. We avoid this issue by forcing all deposits in a mixer to have the same amount. Lastly, Merkle trees are used instead of simpler constructions (e.g., a list) as they provide inclusion proofs, which can later be checked in the ZKP. 

Additional tools and processes can be used to further improve the user experience or increase compliance and anonymity. Examples are transaction relayers \cite{relayer} and viewing keys \cite{viewingKeys}. In this paper, we will not discuss any of these tools, as they are not integral to our protocol and are not used by our prototype. However, there is no reason why our protocol cannot be made compatible with such tools. 

\subsection{Proof of Innocence}
\label{sec:poi}

\begin{figure*}[tb]
    \centering
    \includegraphics[width=0.9\textwidth]{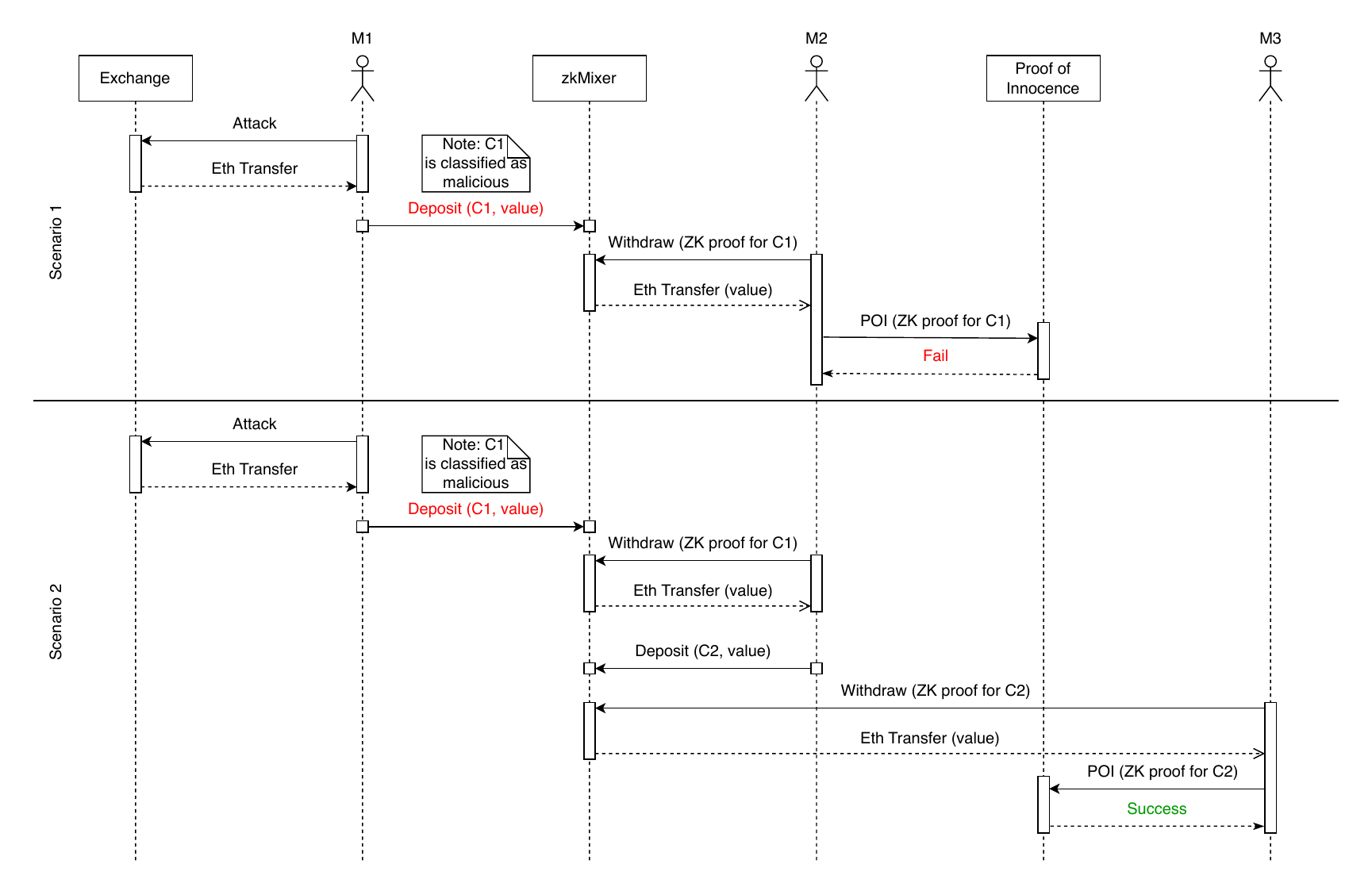}
    \caption{Scenario 1 (top) shows how the Proof of Innocence (PoI) protocol should work: a malicious user, who tries to launder stolen funds, is unable to generate a valid PoI. Scenario 2 (bottom) shows how such a system can be circumvented using a third address and an additional pass through the mixer.}
    \label{fig:example}
\end{figure*}

\noindent
In late 2022, a new method called Proof of Innocence (PoI) was introduced to act as a compliance system for mixers \cite{poi1, poi2}. PoI is an extra voluntary layer on top of a mixer that allows a user to specify a list of existing deposits and then prove that their withdrawal does not correspond to any of them. In practice, these lists are usually compiled by blockchain analytics companies that provide real-time screening services.\footnote{For example, Chainlink provides such a service: \href{https://www.chainalysis.com/free-cryptocurrency-sanctions-screening-tools/}{https://www.chainalysis.com/free-cryptocurrency-sanctions-screening-tools/}} If this list contains all deposits to the mixer that are related to illegal activities, PoI should allow a user to prove that their withdrawal is not linked to proceeds of crime. In contrast, when a malicious user tries to launder funds that come from illicit activities, they will be unable to generate a valid PoI. 
In Fig.~\ref{fig:example}, Scenario 1, we show an example of a correct PoI failure, because the deposit with commitment $C_1$ is included in the list of illicit funds, $L$. 

\subsubsection{Attack on PoI}
PoI does not provide the expected security guarantees, as it is simple to bypass the PoI system by using the mixer itself to first ``clean'' funds before requesting PoI validation. We show an example of PoI bypass in Fig.~\ref{fig:example}, Scenario~2. The process is described in the following. 

\begin{enumerate}
    \item A malicious user, with address $M_1$, hacks an exchange and steals a large amount of Eth
    \item $M_1$ deposits some of their illicit funds to a mixer, creating a deposit with commitment $C_1$
    \item The deposit with $C_1$ is then classified as malicious and added to list $L$, which is used by the PoI system
    \item The malicious user then creates a brand-new address, $M_2$, that is not linked to $M_1$ 
    \item $M_2$ withdraws from the mixer, proving in ZK that they know $C_1$
    \item At this point, $M_2$ would not be able to use PoI, as $C_1$ is in $L$, therefore PoI would fail, as shown in Scenario~1
    \item However, $M_2$'s withdrawal is indistinguishable from any other output from the mixer (as long as PoI is not applied) and so cannot be classified as malicious
    \item So, $M_2$ then sends their new funds to the mixer and creates a new deposit with commitment $C_2$
    \item The deposit with commitment $C_2$ is not added to $L$, as it appears legitimate to outside observers
    \item The malicious user again creates a brand-new address, $M_3$, that is not linked to $M_1$ or $M_2$
    \item $M_3$ then withdraws from the mixer, proving in ZK that they know $C_2$
    \item Finally, $M_3$ successfully generates a valid PoI, as their withdrawal with commitment $C_2$ does not correspond to any of the deposits contained in list $L$
\end{enumerate} 

\noindent
In this way, a malicious user can avoid detection by the PoI system. In addition, they can even claim that their funds are not part of any known malicious activity, which undermines the purpose of the PoI system. The only way to salvage this would be to constantly update the list $L$ every time there is a mixer output; but this would be impractical and self-defeating, as the mixer itself does not trust its own outputs.

\subsubsection{Extensions of PoI}
There are some recent extensions to PoI. Privacy Pools \cite{privacyPools} is a similar protocol to PoI that allows both inclusion and exclusion proofs. Derecho \cite{Derecho} is another similar protocol to PoI that supports proofs for mixers that allow in-pool transactions (also known as ``shielded'' transactions), i.e., transfers of funds that do not leave the mixer. Derecho uses proof-carrying data for recursive proof composition to achieve this. 

However, using Privacy Pools or Derecho does not eliminate the above attack (shown in Fig.~\ref{fig:example}, Scenario~2). In Privacy Pools, every transaction that originates from the mixer would have to be added to the exclusion list unless a proof is made available to all other participants (e.g., published on chain). Similarly, Derecho has no way to track funds that are withdrawn from a mixer, so the same attack can be implemented. 

\subsubsection{PoI in Practice}
RAILGUN\footnote{\url{https://www.railgun.org/}} is a ZK mixer that has implemented PoI \cite{railgun}. However, despite using PoI, Elliptic has identified that around 70\% of the funds that have passed through RAILGUN originate from a single hack \cite{elliptic}. This shows that even without implementing the attack we identified above, PoI systems can be easily bypassed using simple evasive tactics. The reason for this is that RAILGUN uses a static PoI system, which uses a list of malicious addresses provided by a single company (Chainalysis) \cite{poiFail}. That list only includes the addresses sanctioned by OFAC and can miss bad actors, as a sanctioned user can simply move their funds to a different address and then freely use RAILGUN (or any other mixer that relies on static lists). To prevent money laundering, it is clear that more robust techniques must be employed to identify bad actors using mixers. 

\subsection{Research Aim}
\noindent
We have identified that mixers are commonly used by criminals to launder proceeds of crime (Section~\ref{sec:crime}); and we have shown that current PoI systems are insufficient, as they can be easily defeated (Section~\ref{sec:poi}). Therefore, if we want to keep the benefits of anonymity provided by cryptocurrency mixers but also avoid their abuse by bad actors, it is clear that a new design is necessary. To achieve this, we propose a novel design that integrates innocence checks into the mixer (rather than as a separate component, as with current PoI systems); all checks are performed before funds are added to the mixer; and all deposits are verified. This approach ensures that all mixer outputs can be trusted; eliminating the threat of attackers using the mixer to trick the PoI system (i.e., Fig.~\ref{fig:example} Scenario~2).

\section{System Design and Implementation}

\subsection{System Rationale}
\noindent
Existing protocols treat the PoI as a non-compulsory part of the mixer. Therefore, one could attempt to solve the attack vector shown in Fig.~\ref{fig:example} Scenario~2 by integrating the PoI system into the mixer. However, this will not completely solve the issue as a malicious user could deposit funds into the mixer, withdraw the funds (with an invalid PoI), deposit the funds again, and finally withdraw the funds to a third address. In such cases, it is not clear whether the second withdrawal will produce an invalid PoI. If the list used by the PoI system is not updated quickly enough, a valid PoI will be produced. Therefore, to ensure a reliable system, the operator must be able to act quickly, which is difficult to achieve in practice. 

One could attempt to solve this problem by using instantaneous list updates when invalid PoIs are produced. This is possible as the PoI system is now integrated in the mixer's smart contract, so the receiver of a withdrawal with an invalid PoI can be added to the exclusion list of the PoI. However, this does not solve the issue, as a malicious user could first transfer the funds to a new address before performing the second deposit. As the mixer smart contract does not have access to transactions that take place outside the contract, this new address cannot be added to the list automatically. 

One could attempt to mitigate this by making deposits a two-step process to allow for list updates. In this way, when a deposit arrives, it will not be immediately added to the mixer; rather it is held in a staging area for some predefined time to give the operator of the system time to check if the deposit should be added to the PoI list. Then, after the predefined time has elapsed, the user can add their deposit to the mixer. The rest of the mixer operations should work as before.

This approach solves the attack we have identified in Fig.~\ref{fig:example} Scenario~2. However, note that since we have described a system in which all deposits in the mixer can be verified, PoI checks are no longer needed. This is a key idea of our system design, which we introduce in the following sections.

\subsection{Improvements on ZK Mixer Design}\label{sec:imp}
\noindent
Our protocol improves the design of ZK mixers in three main ways. First, we {\em prevalidate the deposits} before they enter the mixer to ensure that they adhere to a set of predefined rules that are configured when the mixer instance is first created. Rules can range from very simple checks that can be performed automatically by smart contracts, such as ``only admit deposits from registered users'', to more complex schemes that check the origin of funds or check a user's historical interactions with other mixers. These latter rules are too complex for a smart contract to perform, and so a group of verifiers is required. The verifiers check every input to the mixer and ensure that it follows the rules of the specific protocol instance. This makes the system highly customizable and eliminates the PoI issue (described earlier in Section~\ref{sec:poi}). 

\begin{figure}[tb]
    \centering
    \includegraphics[width=0.9\linewidth]{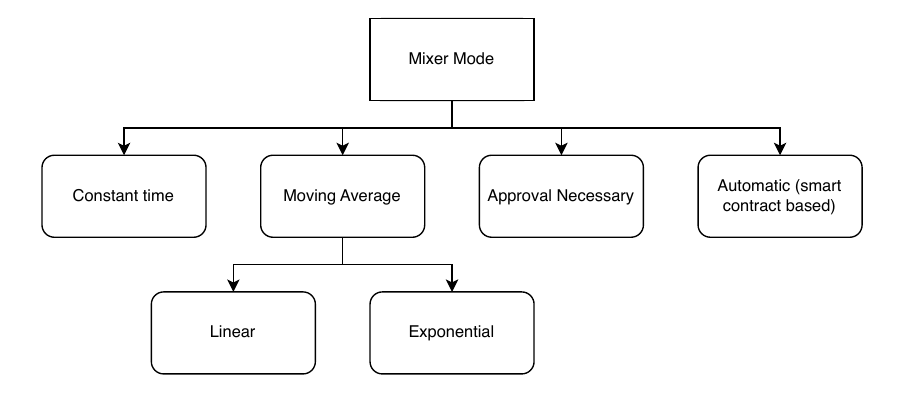}
    \caption{Available configurations for pre-deposit approval modes.}
    \label{fig:delay-modes}
\end{figure}

Second, we introduce a {\em timer delay lock} before tokens are added to the Merkle tree of the mixer. The delay is designed to provide verifiers with enough time to verify each input: it can be set as a constant, e.g., one day; or it can be set to vary dynamically, e.g., increasing with system traffic (see Fig.~\ref{fig:delay-modes} for available delay modes). To be effective, a constant delay must be long enough to ensure that malicious activity can be detected during periods of high traffic; while a variable delay enables both efficient mixer operation during low traffic and sufficient time for validation during high traffic. High traffic bursts over a short period can themselves be an indication of laundering, so dynamically slowing mixer operation during these times acts as a further safety valve. An unusually large number of deposits will trigger a marked increase in the delay time, which flags anomalous deposit behaviour.

Third, we allow verifiers to {\em control inputs to the mixer during the lock period}. Using a common consensus protocol, verifiers have the option of admitting deposits, rejecting and returning deposits to the sender, or confiscating deposits. Although confiscation should be used to restitute funds to victims of crime, it can be abused if the verifiers collude. Therefore, when a system instance is first deployed, the confiscation functionality can be enabled or disabled: when disabled, the verifiers only have the option to admit the deposit or reject and return the deposit to the sender. Here, we must stress again that verifiers can only assume control over deposits {\em during the lock period}. Once a deposit is accepted into the mixer, it is controlled by the user as in any other ZK-based mixer. Many smart contracts commonly used in DeFi have similar, or even stronger, control mechanisms. For example, the issuers of the most popular stablecoins such as USDC, USDT, and BUSD have the power to freeze or burn stablecoins and do so regularly when ordered by authorities \cite{stablecoins}. This property of our system might not satisfy maximalist notions of decentralization, but we believe that it is sufficient for practical use in most cases.

These three additions solve the attack we have identified in Fig.~\ref{fig:example} Scenario~2. The attack relies on the ability to pass funds through the mixer to obfuscate their origin, before passing them through the mixer again to circumvent PoI checks. Unlike previous approaches, where the output of the mixer cannot be trusted, in our proposed approach we can trust that withdrawals follow the rules of the mixer. Therefore, it is not possible to perform the attack.

\subsection{System Components}
\noindent
Our system consists mainly of two smart contracts, the zkMixer and the multiSig. The sequence diagrams in Figs.~\ref{fig:success}-\ref{fig:seize} show the interaction between users and these two smart contracts. Additionally, there are two other auxiliary smart contracts that are used by zkMixer; one that verifies ZKPs and the other that implements the Poseidon hash, as Ethereum does not natively support it. The Poseidon hash \cite{poseidon} was chosen because it is efficient when used in ZKPs.

\subsubsection{Arithmetic Circuits}
\noindent
For the ZK functionality of our protocol we use Circom,\footnote{\url{https://iden3.io/circom}} a programming language used to build arithmetic circuits that can be used in ZKPs; and snarkjs,\footnote{\url{https://github.com/iden3/snarkjs}} a library implementing zkSNARK schemes. The arithmetic circuits are first compiled using Circom, and then snarkjs is used to create a smart contract that can verify the ZKPs for withdrawals on-chain. For our prototype, the Groth16 scheme was used as it is the most popular, but this can be easily substituted by other schemes, such as PlonK. 

Using Groth16 requires a circuit-specific trusted setup and a one-time procedure to generate public parameters, which also generates an unwanted trapdoor as a byproduct \cite{trusted_setup}. This trapdoor can be used to create fraudulent proofs; therefore users must trust that this procedure is done in a way that no single party has control over the trapdoor or that the trapdoor was discarded by the creator of the system. Groth16 could be substituted with a different proving system (such as PlonK) that only requires a universal trusted setup. Several universal trusted setup ceremonies have been performed by others using secure multiparty computation (e.g., Semaphore),\footnote{\url{https://github.com/privacy-scaling-explorations/perpetualpowersoftau}} which eliminates trust in the creator of this system.

We have implemented three arithmetic circuits. The first circuit creates the commitment from the secret and the nullifier by applying the Poseidon Hash. The second circuit, uses the first and is responsible for calculating the root of the Merkle tree given a commitment, and the rest of the tree nodes in the path from the commitment to the root. Finally, the last circuit, which uses the previous two, is responsible for the ZKPs that are used to withdraw from the mixer. This circuit takes the user's input, calculates the user's commitment, calculates the root of the Merkle tree, and finally verifies that the root is the same as the root in the smart contract. 

\subsubsection{zkMixer Smart Contract}
\noindent
The zkMixer smart contract contains the core functionality of our mixer. It follows the design presented in Section~\ref{sec:ZKMixer}, with some additional functionality as described in Section~\ref{sec:imp}. zkMixer contains a list of pre-deposits, i.e., deposits before they are added to the mixer's Merkle tree, along with functionality to freeze and unfreeze them, to add them to the mixer, and to delete them and send the funds to a different address (either refund or confiscate them). The properties of the mixer, such as the mixer mode and the deposit amount, must be configured prior to deployment. As shown in Fig.~\ref{fig:delay-modes}, the mixer operates in one of four distinct modes, based on how pre-deposits are approved. The modes are as follows:
\begin{enumerate}
    \item Constant time: Deposits have a constant lock time. The lock time can be set according to the capabilities of the verifiers.
    \item Moving average: Deposits have a variable lock time that depends on the moving average of mixer traffic. The lock time can also be configured to increase linearly or exponentially when the deposit count exceeds the moving average. The minimum lock time and other constants that influence the moving average can be set to satisfy the requirements of each use case.
    \item Approval necessary: In this mode, deposits will be added only after the verifiers approve them.
    \item Automatic: This mode does not use verifiers or lock time, but instead deposits are added to the mixer if they satisfy some conditions. The exact conditions can be configured by the user under the requirement that they must be verifiable by a smart contract.
\end{enumerate}

\subsubsection{Variable Lock Time}
When the mixer mode is set to moving average, deposits will be delayed for some variable time period. We first calculate a moving average of mixer traffic, over some specified window, to capture normal deposit behaviour. We then check if the current period is experiencing increased traffic and, if so, adjust the system delay. There are two delay modes; one uses a linear delay function, and the other an exponential function. 
For linear mode, the delay $D$ is calculated as,

\begin{equation} \label{eq:linear-mode}
D = 
\begin{cases}
(C-A) \times M, & \text{if } C > A\\
M,              & \text{otherwise},
\end{cases}
\end{equation}

\noindent
and for exponential mode,

\begin{equation} \label{eq:variable-mode}
D = 
\begin{cases}
2^{C-A} \times M, & \text{if } C > A\\
M,              & \text{otherwise},
\end{cases}
\end{equation}

\noindent
where $C$ is the running count of deposits during the current period, $A$ is the current moving average, and $M$ is the minimum delay of the system.

\subsubsection{multiSig Smart Contract}
The multiSig smart contract contains all the functionality used by the verifiers. The smart contract has a manager who can add and remove verifiers. This can be a centralized authority, a DAO, or something similar depending on the intended use case. Verifiers are responsible for checking that the deposits conform to the rules of the mixer. 
These verifiers can be users of the mixer that want to participate in its governance or blockchain analytics companies that provide real-time screening services.\footnote{Examples include Elliptic (\url{https://www.elliptic.co/solutions/screening}), TRM (\url{https://www.trmlabs.com/blockchain-intelligence-platform/wallet-screening}), and Chainalysis (\url{https://www.chainalysis.com/solution/crypto-compliance-assess/})} 
The services of the verifiers can be rewarded by adding a fee to all deposits, although for simplicity this has not been implemented in our prototype. More advanced incentive and punishment mechanisms can also be implemented, but again for simplicity these were omitted from our prototype.

\begin{figure}[tb]
    \centering
    \includegraphics[width=0.8\linewidth]{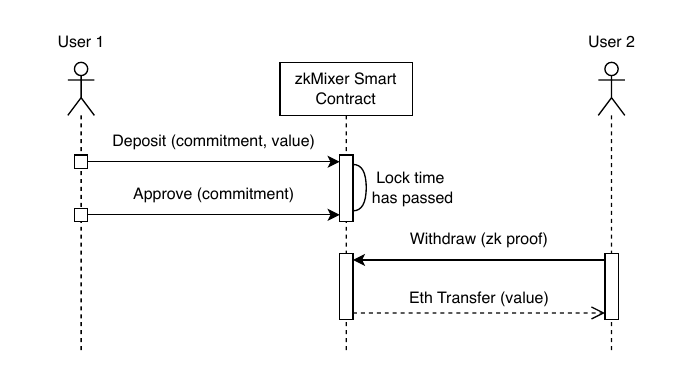}
    \caption{zkMixer use case: showing a successful deposit that is not flagged. Users can approve their own deposits after the lock time has passed.}
    \label{fig:success}
\end{figure}

\begin{figure}[tb]
    \centering
    \includegraphics[width=0.95\linewidth]{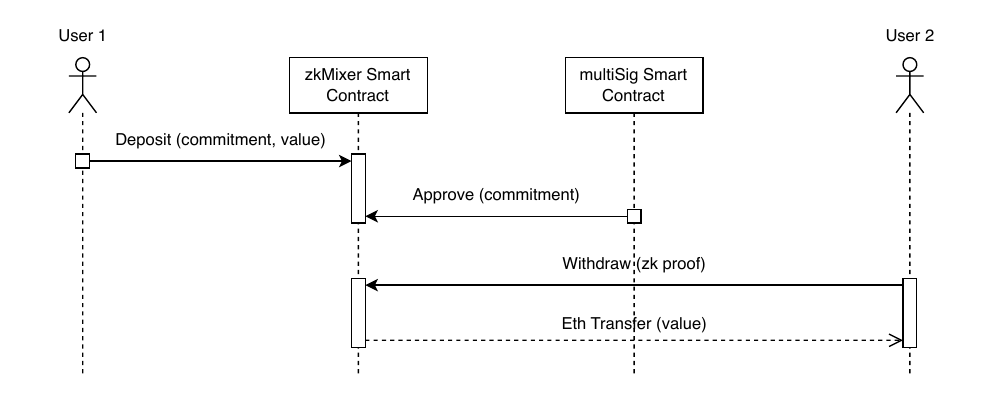}
    \caption{zkMixer use case: showing a successful deposit, when approval is by consensus  using the multiSig smart contract.}
    \label{fig:success2}
\end{figure}

Any single verifier using the multiSig contract can freeze any of the pre-deposits to the mixer. Freezing a deposit indicates that they have potentially identified a deposit that does not conform to the requirements of the mixer. At that point, the deposit cannot be added to the mixer even if its lock time has elapsed. A verifier can then propose what should be done with the deposit and indicate an address to which the deposit should be sent. If the multiSig smart contract was configured to allow for confiscation of funds, the address can be any address, but if it was not, the address is set to the address of the user who deposited the funds, i.e., only refunds are allowed. The rest of the verifiers then vote on the proposal. If the majority vote for the proposal, then the pre-deposit is deleted from the mixer, and deposited funds are sent to the specified address. However, if the majority of the verifiers do not agree with the proposal, then the proposal fails and the deposit is admitted to the mixer. MultiSig approvals work similarly.

\subsection{Use cases}

\noindent
Here, we present a series of UML sequence diagrams to elaborate common zkMixer use cases.

Fig.~\ref{fig:success} and Fig.~\ref{fig:success2} show two examples of successful deposits. In Fig.~\ref{fig:success}, the user first makes a deposit to the mixer, which creates a pre-deposit on zkMixer. Then, after the lock time has passed, as the pre-deposit was not flagged, the user is allowed to call the approval function, which deletes the pre-deposit and adds the user's commitment to the mixer's Merkle tree. Some time later, the funds are withdrawn anonymously from the mixer by a different user. Fig.~\ref{fig:success2} shows a different example of successful use of the mixer. The user again makes a deposit, but in this case the verifiers use multiSig to approve the deposit by consensus, before a withdrawal takes place.

\begin{figure}[tb]
    \centering
    \includegraphics[width=0.9\linewidth]{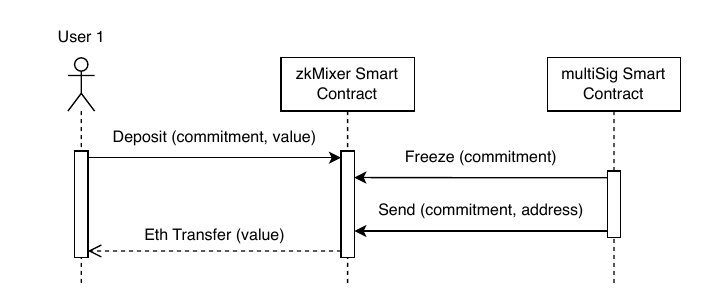}
    \caption{zkMixer use case: showing a deposit that is rejected through consensus. The deposit is first frozen and then returned to the sender.}
    \label{fig:refund}
\end{figure}

\begin{figure}[tb]
    \centering
    \includegraphics[width=0.95\linewidth]{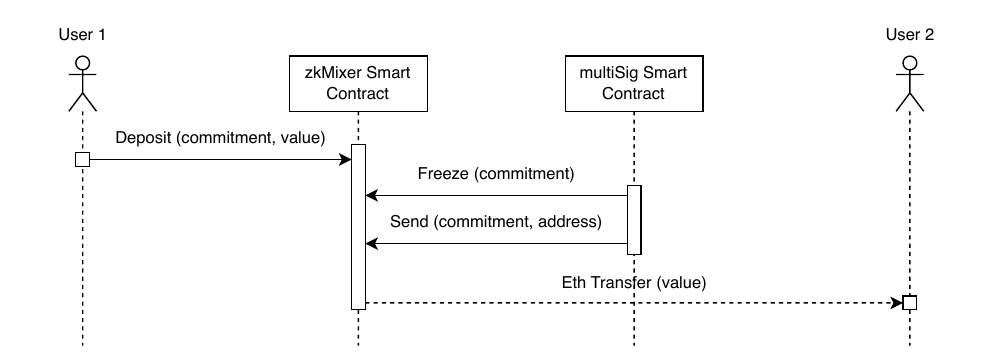}
    \caption{zkMixer use case: showing a deposit that is confiscated through consensus after User 1 is identified as laundering funds. After freezing, the deposit is sent to the rightful owner of the funds, User 2.}
    \label{fig:seize}
\end{figure}

Examples of failure cases are shown in Fig.~\ref{fig:refund} and Fig.~\ref{fig:seize}. The former shows an example of a refund, while the latter shows an example of a confiscation. In both cases, the user makes a deposit to the mixer, which is found suspicious by a verifier and frozen using the multiSig. In the first example (Fig.~\ref{fig:refund}), the verifier makes a proposal to reject the deposit and refund the user. The rest of the verifiers reach consensus and approve the proposal. The funds are then transferred to the original sender, and the deposit is removed from the smart contract. In the second example (Fig.~\ref{fig:seize}), the verifier proposes a confiscation, which is approved by consensus. In this case, the funds are sent to a different user, who is known to be the rightful owner of the stolen funds.

\section{Analysis and Discussion}\label{sec:analysis}

\subsection{Costs of Deployment and Operations}
\noindent
We have analysed the cost of deploying and running an instance of zkMixer. 
Table~\ref{tab:gas1} shows the cost of deployment, which is performed only once per mixer instance. Using current gas and ETH prices of 1.00 Gwei and \$2000, respectively, the total deployment cost is only \$21.22. Likewise, from Table~\ref{tab:gas2}, we also see that the cost of operation is very low, with all operations costing less than one dollar each. Furthermore, since our prototype is not optimized for performance and efficiency, these costs are likely to decrease.

Note that in our prototype, the mixers have no fees. However, in a realistic application, fees would be needed to compensate the network of verifiers, so the cost of using a mixer will be higher than presented.

\begin{table}[bt!]
\centering
\caption{Gas (Gwei) and Monetary (USD) Cost of Contract Deployment}\label{tab:gas1}
\resizebox{0.6\linewidth}{!}{%
\begin{tabular}{l r r}
\toprule
Smart Contract & Gas Cost & \$ Cost\\
\midrule
Groth16Verifier & 428,190 & \$0.86 \\
multiSig & 1,669,860 & \$3.34 \\
zkMixer  & 3,265,751 & \$6.53 \\
PoseidonT3 & 5,245,573 & \$10.49 \\
\bottomrule
\end{tabular}
} 
\end{table}

\begin{table}[bt!]
\centering
\caption{Gas (Gwei) and Monetary (USD) Cost of Operations}\label{tab:gas2}
\resizebox{0.9\linewidth}{!}{%
\begin{tabular}{c l r c}
\toprule
Smart Contract & Operation & Gas Cost & \$ Cost\\ 
\midrule
zkMixer & Deposit & 102,299 & \$0.20 \\
zkMixer & Withdraw & 279,198 & \$0.56 \\
zkMixer & Approve & 452,440 & \$0.90 \\
\midrule
multiSig & Remove Verifier & 31,106 & \$0.06 \\
multiSig & Tally & 50,295 & \$0.10 \\
multiSig & Vote & 52,521 & \$0.11 \\
multiSig & Add Verifier & 53,039 & \$0.11 \\
multiSig & Propose & 154,033 & \$0.31 \\
\bottomrule
\end{tabular}
}
\end{table}

\subsection{Lock Time}\label{sec:lock}
\noindent
The lock time is one of the most important properties of our protocol. Many variables, such as the capabilities of the verifiers and the expected traffic to the mixer, must be considered before an appropriate lock time is selected. In general, for safety, higher lock times should be preferred as they allow more time for verification. If lock times are too short, it can lead to verifiers being overwhelmed and incapable of adequately scrutinizing deposits. This opens a potential exploit for malicious actors, where large amounts of deposits are deliberately submitted in quick succession, with the aim of causing verifiers to have insufficient time to vet all deposits before the lock time runs out.

Our protocol has two measures to combat this. First, there is the variable lock time; mixers that use this mode will have more time to examine the deposits when the traffic is high. Second, verifiers could decide to freeze all deposits at periods of high activity if they feel that there is insufficient capacity to check them all within the lock time. Deposits can then be examined without time pressure until the backlog is cleared.  Currently, this functionality can only be performed manually, meaning that a verifier must freeze each incoming deposit individually; but the protocol can easily be modified to add a cut-off threshold such that any subsequent deposits will be automatically frozen until the number of deposits returns to normal levels.

Longer lock times could potentially make the system less user-friendly. That being said, in order to prevent deanonymization, mixer users should ideally retain their deposits in the mixer for some time before withdrawing them. Tornado Cash, in a blog post, recommended that its users wait for at least five additional deposits after their deposit, and also for at least 24 hours, before withdrawing \cite{tornadoBlog}. The rationale is that if there is a deposit followed immediately by a withdrawal, then an observer could guess that the two originate from the same user. Therefore, introducing a lock time should not negatively affect the usability of the mixer.

\subsection{Network of Verifiers}
\noindent
The current prototype uses a consensus mechanism implemented by a multiSig smart contract. This is simple and provides the desirable functionality, but might not be ideal for a real-world deployment. Such a system requires incentives for participants and thus could be very expensive in practice. In addition, it introduces a trust source and is susceptible to collusion. The manager and verifiers have too much control over the system and could act maliciously, either denying user access or even stealing funds. An alternative solution, such as Decentralized Oracle Networks (DONs), could be used instead. The oracles in DONs have the necessary knowledge and experience to run mission-critical infrastructure, and are trusted to act as the backbone of many DeFi protocols. 

In any case, the verifiers could be existing transaction screening service providers. These providers already identify and track fraudulent activity on chain and so could easily monitor the deposits to zkMixer. Real-time AI-based scoring systems could also be used to provide a risk score for each deposit \cite{privacyPools}.

\subsection{Deposit Verification}
\noindent
We assume that the network of verifiers will be able to identify fraudulent deposits to the mixer. This is a difficult task, and if the verifiers are not up to it, then the whole protocol fails. We believe that verification can be achieved using a combination of self-reporting from exploited parties and dedicated nodes running tracing programs. When a hack is noticed, the affected party can report it to a registry. An example of such reporting is done by Bybit for the recent hack they suffered.\footnote{\url{https://www.lazarusbounty.com/en/}} Then the nodes will be able to maintain a banned list of all the addresses that the funds move through. If any of those addresses attempts to interact with the mixer, it can be flagged by the verifiers.

\subsection{Interactions between Mixers}
\noindent
This protocol can be configured to suit the needs of different users and businesses. We imagine that this will lead to many instances of mixers operating under different rule sets. For example, a mixer operating in the US might base its rules on the OFAC list of sanctioned addresses, while one operating in Russia might use a different list. This creates a problem as different rule sets might be incompatible, resulting in a situation where different mixers may reject outputs from each other; e.g., a user who has previously received funds from the Russian mixer might be unable to deposit into the US mixer. 

We recognize this problem, but we believe that our protocol provides a solution. In our example, a third mixer (which we will call ``High Bar'') can be set up to require verification of all deposits, and there will be very high requirements on deposits before they are allowed. The High Bar mixer will be operated by verifiers who can be contacted by users off-chain and who are willing to spend time auditing the source of the user's funds (such verifiers could be a government department or national regulator, for instance).\footnote{Similar to applications to OFAC for licenses to  otherwise prohibited transactions (\url{https://ofac.treasury.gov/ofac-license-application-page})} If the audit shows that the funds received from the Russian mixer are legitimate and do not originate from illicit activities, then the user will be able to pass these funds through the High Bar mixer. Due to the increased level of scrutiny provided by High Bar, all other US-based mixers will accept its output as legitimate funds.

\subsection{Fracturing of Anonymity Pools}
\noindent
One of the benefits of existing mixers is the large anonymity sets. In existing mixers, the pools of funds are usually only broken to allow for different deposit amounts. For example, Tornado Cash had four pools corresponding to 0.1, 1, 10, and 100 Eth. In our protocol, the pools of funds are not only broken to allow for different amounts; pools are also based on user requirements, legislation, use cases, etc. This comes at the expense of fracturing the anonymity set of users, thus reducing the privacy protection of the system. 

We recognize that this is an issue, but we believe that in practice only a handful of instances of the mixer will be sufficient. For example, a mixer based on the OFAC list will be suitable not only for the US but for most jurisdictions and could act as a baseline mixer that most people would be willing to accept. Some use cases, for example financial institutions, might have higher levels of requirements such as know-your-customer (KYC) rules. However, again, most financial institutions will require similar levels of certainty, and so a High Bar mixer that satisfies the highest of those could be set up; then, all institutions could ask users to use that High Bar mixer to ensure that their own requirements are either satisfied or exceeded. 

\subsection{Concrete Instantiation}
\label{sec:ci}
\noindent
For completeness, we include a practical, concrete example of how the protocol could be deployed. {\bf Lock time}: we believe that a variable lock time is best for managing periods of high traffic. {\bf Non-confiscation}: we encourage against allowing verifiers to confiscate deposits. Confiscation not only opens the protocol to an additional attack vector, but also limits its reach, as many users will avoid it due to the lack of custody during deposits. {\bf Verifiers}: we suggest that the system has a small number of verifiers (e.g., 3), each of which should be an established screening company. This not only reduces fees, but, as the verifiers cannot confiscate deposits, it does not severely impact security. {\bf Funds tracking}: finally, we would expect such a protocol to only track funds from relatively large hacks or thefts, as it is much more difficult for verifiers to adjudicate on small instances of theft, such as a dispute between two users. This, along with self-reporting, can make tracking feasible. 

In Table~\ref{tab:comp}, we compare our concrete instantiation against a ZK mixer using PoI. As can be seen, the concrete instantiation is slower, has higher fees, and is susceptible to denial-of-service (DoS) attacks if the validators are malicious. However, the concrete instantiation provides better compliance as it avoids issues with PoI, and instead uses transaction screening to prevent bad actors from using the system. As explained in Section~\ref{sec:lock}, speed is not critical for a mixer. Using only a few verifiers should also help minimize the additional cost. Therefore, DoS is the only major drawback of this concrete instantiation; but as all interactions are public on chain, it would be easy for users to prove that DoS is taking place. Overall, we believe that this is a worthwhile drawback given the compliance benefits.

\begin{table}[t]
\centering
\caption{Comparison between concrete instantiation (described in Section~\ref{sec:ci}) and a ZK mixer using PoI}
\label{tab:comp}
\resizebox{0.9\linewidth}{!}{%
\begin{tabular}{l|l l}
\toprule
\backslashbox{Properties}{Protocol} & Concrete Instantiation & ZK Mixer using PoI \\
\midrule
Speed & Initial lock time & Instantaneous\\
Cost & Gas and validator fees & Only gas fees \\
Trust & Denial of serivce possible & Zero trust \\
Compliance & Deposit screening & PoI can be circumvented \\
\bottomrule
\end{tabular}
} 
\end{table}

\section{Conclusion} 
\noindent
In this paper, we examined cryptocurrency mixers, we categorized them based on their characteristics, we detailed how they operate, and we presented evidence of their use in money laundering. We then discussed Proof of Innocence (PoI), a solution previously proposed to prevent money laundering in mixers, and identified why PoI cannot work in practice. To address these issues, we introduced a novel mixer protocol that mitigates the PoI problem by using a consensus mechanism. Our protocol design allows groups of users to set up different instances of the mixer, each governed by the requirements that the group wants to achieve. We release a full prototype of our protocol, which we have made open-source available at \ifnum\ANON=1 [redacted]\else \url{https://github.com/lifeisbeer/zkMixer}\fi.

\ifnum\ANON=1
\else
    \ifCLASSOPTIONcompsoc
      \section*{Acknowledgments}
    \else
      \section*{Acknowledgment}
    \fi
    \noindent
    This work was supported by UKRI EPSRC Grant No. EP/Y028392/1: AI for Collective Intelligence (AI4CI). The authors have no conflicts of interest to declare. 
\fi



\bibliographystyle{IEEEtran}
\bibliography{bibliography.bib}

\begin{thebibliography}{10}
\providecommand{\url}[1]{#1}
\csname url@samestyle\endcsname
\providecommand{\newblock}{\relax}
\providecommand{\bibinfo}[2]{#2}
\providecommand{\BIBentrySTDinterwordspacing}{\spaceskip=0pt\relax}
\providecommand{\BIBentryALTinterwordstretchfactor}{4}
\providecommand{\BIBentryALTinterwordspacing}{\spaceskip=\fontdimen2\font plus
\BIBentryALTinterwordstretchfactor\fontdimen3\font minus \fontdimen4\font\relax}
\providecommand{\BIBforeignlanguage}[2]{{%
\expandafter\ifx\csname l@#1\endcsname\relax
\typeout{** WARNING: IEEEtran.bst: No hyphenation pattern has been}%
\typeout{** loaded for the language `#1'. Using the pattern for}%
\typeout{** the default language instead.}%
\else
\language=\csname l@#1\endcsname
\fi
#2}}
\providecommand{\BIBdecl}{\relax}
\BIBdecl

\bibitem{monero}
\BIBentryALTinterwordspacing
N.~{van Saberhagen}, ``Cryptonote v 2.0,'' Monero, Tech. Rep., 2013. [Online]. Available: \url{https://www.getmonero.org/resources/research-lab/pubs/cryptonote-whitepaper.pdf}
\BIBentrySTDinterwordspacing

\bibitem{zerocash}
\BIBentryALTinterwordspacing
E.~Ben~Sasson, A.~Chiesa, C.~Garman, M.~Green, I.~Miers, E.~Tromer, and M.~Virza, ``Zerocash: Decentralized anonymous payments from {Bitcoin},'' in \emph{2014 IEEE Symposium on Security and Privacy}, 2014. [Online]. Available: \url{https://doi.org/10.1109/SP.2014.36}
\BIBentrySTDinterwordspacing

\bibitem{bydit}
\BIBentryALTinterwordspacing
Elliptic, ``The largest theft in history - following the money trail from the {B}ybit {H}ack,'' 2025, accessed: 2025-03-10. [Online]. Available: \url{https://www.elliptic.co/blog/bybit-hack-largest-in-history}
\BIBentrySTDinterwordspacing

\bibitem{2025report}
\BIBentryALTinterwordspacing
Chainalysis, ``The 2025 crypto crime report,'' Chainalysis, Tech. Rep., 2025. [Online]. Available: \url{https://www.chainalysis.com/wp-content/uploads/2025/03/the-2025-crypto-crime-report-release.pdf}
\BIBentrySTDinterwordspacing

\bibitem{2024report}
\BIBentryALTinterwordspacing
------, ``The 2024 crypto crime report,'' Chainalysis, Tech. Rep., 2024. [Online]. Available: \url{https://www.chainalysis.com/wp-content/uploads/2024/06/the-2024-crypto-crime-report-release.pdf}
\BIBentrySTDinterwordspacing

\bibitem{mixerUsage}
\BIBentryALTinterwordspacing
------, ``Crypto money laundering: Four exchange deposit addresses received over \$1 billion in illicit funds in 2022,'' 2023, accessed: 2025-03-09. [Online]. Available: \url{https://www.chainalysis.com/blog/crypto-money-laundering-2022/}
\BIBentrySTDinterwordspacing

\bibitem{dusting}
\BIBentryALTinterwordspacing
L.~Wright and O.~Adejumo, ``{DeFi} protocols {A}ave, {U}niswap, {B}alancer, ban users following {OFAC} sanctions on {T}ornado {C}ash,'' 2022. [Online]. Available: \url{https://cryptoslate.com/defi-protocols-aave-uniswap-balancer-ban-users-following-ofac-sanctions-on-tornado-cash/}
\BIBentrySTDinterwordspacing

\bibitem{bestmixer}
\BIBentryALTinterwordspacing
Europol, ``Multi-million euro cryptocurrency laundering service {B}estmixer.io taken down,'' 2019. [Online]. Available: \url{https://www.europol.europa.eu/media-press/newsroom/news/multi-million-euro-cryptocurrency-laundering-service-bestmixerio-taken-down}
\BIBentrySTDinterwordspacing

\bibitem{chip}
\BIBentryALTinterwordspacing
DOJ, ``Justice department investigation leads to takedown of darknet cryptocurrency mixer that processed over \$3 billion of unlawful transactions,'' 2023, accessed: 2025-03-09. [Online]. Available: \url{https://www.justice.gov/archives/opa/pr/justice-department-investigation-leads-takedown-darknet-cryptocurrency-mixer-processed-over-3}
\BIBentrySTDinterwordspacing

\bibitem{sinbad}
\BIBentryALTinterwordspacing
FIOD, ``{FIOD} takes large crypto currency mixer off the air,'' 2023, accessed: 2025-03-09. [Online]. Available: \url{https://www.fiod.nl/fiod-takes-large-crypto-currency-mixer-off-the-air/}
\BIBentrySTDinterwordspacing

\bibitem{fog}
\BIBentryALTinterwordspacing
DOJ, ``Bitcoin {F}og operator convicted of money laundering conspiracy,'' 2024. [Online]. Available: \url{https://www.justice.gov/archives/opa/pr/bitcoin-fog-operator-convicted-money-laundering-conspiracy}
\BIBentrySTDinterwordspacing

\bibitem{helix}
\BIBentryALTinterwordspacing
------, ``Operator of {H}elix darknet cryptocurrency “mixer” sentenced in money laundering conspiracy and ordered to forfeit over \$400m in assets,'' 2024, accessed: 2025-03-09. [Online]. Available: \url{https://www.justice.gov/archives/opa/pr/operator-helix-darknet-cryptocurrency-mixer-sentenced-money-laundering-conspiracy-and}
\BIBentrySTDinterwordspacing

\bibitem{samourai}
\BIBentryALTinterwordspacing
------, ``Founders and {CEO} of cryptocurrency mixing service arrested and charged with money laundering and unlicensed money transmitting offenses,'' 2024, accessed: 2025-03-09. [Online]. Available: \url{https://www.justice.gov/usao-sdny/pr/founders-and-ceo-cryptocurrency-mixing-service-arrested-and-charged-money-laundering}
\BIBentrySTDinterwordspacing

\bibitem{tornadoArrest1}
\BIBentryALTinterwordspacing
------, ``Tornado {C}ash founders charged with money laundering and sanctions violations,'' 2023, accessed: 2025-03-09. [Online]. Available: \url{https://www.justice.gov/usao-sdny/pr/tornado-cash-founders-charged-money-laundering-and-sanctions-violations}
\BIBentrySTDinterwordspacing

\bibitem{blender}
\BIBentryALTinterwordspacing
OFAC, ``{U.S.} {T}reasury issues first-ever sanctions on a virtual currency mixer, targets {DPRK} cyber threats,'' 2022. [Online]. Available: \url{https://home.treasury.gov/news/press-releases/jy0768}
\BIBentrySTDinterwordspacing

\bibitem{tornadoSanction1}
\BIBentryALTinterwordspacing
------, ``{U.S.} {T}reasury sanctions notorious virtual currency mixer {T}ornado {C}ash,'' 2022, accessed: 2025-03-09. [Online]. Available: \url{https://home.treasury.gov/news/press-releases/jy0916}
\BIBentrySTDinterwordspacing

\bibitem{sinbadSanction}
\BIBentryALTinterwordspacing
------, ``Counter terrorism designations; {I}ran-related designations; cyber-related designation; {N}orth {K}orea designation,'' 2023. [Online]. Available: \url{https://ofac.treasury.gov/recent-actions/20231129}
\BIBentrySTDinterwordspacing

\bibitem{tornadoAppeal}
\BIBentryALTinterwordspacing
D.~R. Willett, ``Van loon et al. v. {D}epartment of the {T}reasury,'' 2024, accessed: 2025-03-09. [Online]. Available: \url{https://www.ca5.uscourts.gov/opinions/pub/23/23-50669-CV0.pdf}
\BIBentrySTDinterwordspacing

\bibitem{coinjoin1}
\BIBentryALTinterwordspacing
G.~Maxwell, ``I taint rich!'' accessed: 2025-03-07. [Online]. Available: \url{https://bitcointalk.org/index.php?topic=139581}
\BIBentrySTDinterwordspacing

\bibitem{coinjoin2}
\BIBentryALTinterwordspacing
------, ``Coin{J}oin: {B}itcoin privacy for the real world,'' accessed: 2025-03-07. [Online]. Available: \url{https://bitcointalk.org/index.php?topic=279249}
\BIBentrySTDinterwordspacing

\bibitem{ZKP}
\BIBentryALTinterwordspacing
S.~Goldwasser, S.~Micali, and C.~Rackoff, ``The knowledge complexity of interactive proof systems,'' \emph{SIAM Journal on Computing}, vol.~18, no.~1, pp. 186--208, 1989. [Online]. Available: \url{https://doi.org/10.1137/0218012}
\BIBentrySTDinterwordspacing

\bibitem{blind}
\BIBentryALTinterwordspacing
D.~Chaum, ``Blind signatures for untraceable payments,'' in \emph{Advances in Cryptology}.\hskip 1em plus 0.5em minus 0.4em\relax Boston, MA: Springer US, 1983, pp. 199--203. [Online]. Available: \url{https://doi.org/10.1007/978-1-4757-0602-4_18}
\BIBentrySTDinterwordspacing

\bibitem{coinShuffle}
\BIBentryALTinterwordspacing
T.~Ruffing, P.~Moreno-Sanchez, and A.~Kate, ``{CoinShuffle}: Practical decentralized coin mixing for {Bitcoin},'' in \emph{Computer Security - ESORICS 2014}, 2014, pp. 345--364. [Online]. Available: \url{https://link.springer.com/chapter/10.1007/978-3-319-11212-1_20}
\BIBentrySTDinterwordspacing

\bibitem{zkSnark}
\BIBentryALTinterwordspacing
N.~Bitansky, R.~Canetti, A.~Chiesa, and E.~Tromer, ``From extractable collision resistance to succinct non-interactive arguments of knowledge, and back again,'' in \emph{Proceedings of the 3rd Innovations in Theoretical Computer Science Conference}, ser. ITCS '12, 2012, p. 326–349. [Online]. Available: \url{https://doi.org/10.1145/2090236.2090263}
\BIBentrySTDinterwordspacing

\bibitem{zkSnarkSurvey}
\BIBentryALTinterwordspacing
A.~Nitulescu, ``{zk-SNARKs}: A gentle introduction,'' {\'E}cole Normale Sup{\'e}rieure, Tech. Rep., 2020. [Online]. Available: \url{https://www.di.ens.fr/~nitulesc/files/Survey-SNARKs.pdf}
\BIBentrySTDinterwordspacing

\bibitem{Pinocchio}
\BIBentryALTinterwordspacing
B.~Parno, J.~Howell, C.~Gentry, and M.~Raykova, ``Pinocchio: Nearly practical verifiable computation,'' in \emph{2013 IEEE Symposium on Security and Privacy}, 2013, pp. 238--252. [Online]. Available: \url{https://doi.org/10.1145/2856449}
\BIBentrySTDinterwordspacing

\bibitem{groth16}
\BIBentryALTinterwordspacing
J.~Groth, ``On the size of pairing-based non-interactive arguments,'' in \emph{Advances in Cryptology -- EUROCRYPT 2016}, 2016, pp. 305--326. [Online]. Available: \url{http://dx.doi.org/10.1007/978-3-662-49896-5 11}
\BIBentrySTDinterwordspacing

\bibitem{PLONK}
\BIBentryALTinterwordspacing
A.~Gabizon, Z.~J. Williamson, and O.~Ciobotaru, ``{PLONK}: Permutations over lagrange-bases for oecumenical noninteractive arguments of knowledge,'' Cryptology {ePrint} Archive, Paper 2019/953, 2019. [Online]. Available: \url{https://eprint.iacr.org/2019/953}
\BIBentrySTDinterwordspacing

\bibitem{zkSnarkSurvey2}
\BIBentryALTinterwordspacing
J.~Liang, D.~Hu, P.~Wu, Y.~Yang, Q.~Shen, and Z.~Wu, ``{SoK}: Understanding zk-{SNARKs}: The gap between research and practice,'' Cryptology {ePrint} Archive, Paper 2025/172, 2025. [Online]. Available: \url{https://eprint.iacr.org/2025/172}
\BIBentrySTDinterwordspacing

\bibitem{tornado}
\BIBentryALTinterwordspacing
A.~Pertsev, R.~Semenov, and R.~Storm, ``Tornado {C}ash privacy solution version 1.4,'' Tornado {C}ash, Tech. Rep., 2019. [Online]. Available: \url{https://berkeley-defi.github.io/assets/material/Tornado%20Cash%20Whitepaper.pdf}
\BIBentrySTDinterwordspacing

\bibitem{MerkleTree}
\BIBentryALTinterwordspacing
R.~C. Merkle, ``{Secrecy, Authentication, and Public Key Systems.}'' Ph.D. dissertation, Stanford University, Stanford, CA, USA, 1979. [Online]. Available: \url{https://www.ralphmerkle.com/papers/Thesis1979.pdf}
\BIBentrySTDinterwordspacing

\bibitem{relayer}
\BIBentryALTinterwordspacing
I.~A. Seres, ``On blockchain metatransactions,'' in \emph{2020 IEEE International Conference on Blockchain (Blockchain)}, 2020. [Online]. Available: \url{https://doi.org/10.1109/Blockchain50366.2020.00029}
\BIBentrySTDinterwordspacing

\bibitem{viewingKeys}
\BIBentryALTinterwordspacing
{Electric Coin Company}, ``Explaining viewing keys,'' 2020. [Online]. Available: \url{https://electriccoin.co/blog/explaining-viewing-keys/}
\BIBentrySTDinterwordspacing

\bibitem{poi1}
\BIBentryALTinterwordspacing
Chainway, ``Proof of innocence built on {T}ornado {C}ash ({GitHub} repository),'' 2022, accessed: 2025-03-09. [Online]. Available: \url{https://github.com/chainwayxyz/proof-of-innocence}
\BIBentrySTDinterwordspacing

\bibitem{poi2}
\BIBentryALTinterwordspacing
------, ``Introducing proof of innocence built on {T}ornado {C}ash,'' 2023, accessed: 2025-03-09. [Online]. Available: \url{https://medium.com/@chainway_xyz/introducing-proof-of-innocence-built-on-tornado-cash-7336d185cda6}
\BIBentrySTDinterwordspacing

\bibitem{privacyPools}
\BIBentryALTinterwordspacing
V.~Buterin, J.~Illum, M.~Nadler, F.~Schär, and A.~Soleimani, ``Blockchain privacy and regulatory compliance: Towards a practical equilibrium,'' \emph{Blockchain: Research and Applications}, vol.~5, no.~1, p. 100176, 2024. [Online]. Available: \url{https://www.sciencedirect.com/science/article/pii/S2096720923000519}
\BIBentrySTDinterwordspacing

\bibitem{Derecho}
\BIBentryALTinterwordspacing
J.~Beal and B.~Fisch, ``Derecho: Privacy pools with proof-carrying disclosures,'' in \emph{Proceedings of the 2024 on ACM SIGSAC Conference on Computer and Communications Security}, 2024, p. 3197–3211. [Online]. Available: \url{https://doi.org/10.1145/3658644.3670270}
\BIBentrySTDinterwordspacing

\bibitem{railgun}
\BIBentryALTinterwordspacing
RAILGUN, ``Having your privacy \& eating it too — {RAILGUN} private proofs of innocence,'' 2023, accessed: 2025-03-09. [Online]. Available: \url{https://medium.com/@Railgun_Project/having-your-privacy-eating-it-too-railgun-proof-of-innocence-efcba557aac4}
\BIBentrySTDinterwordspacing

\bibitem{elliptic}
\BIBentryALTinterwordspacing
Elliptic, ``{FBI} confirms {N}orth {K}orea’s {L}azarus {G}roup as hackers behind \$100 million {H}armony {H}orizon {B}ridge theft,'' 2024, accessed: 2025-03-09. [Online]. Available: \url{https://www.elliptic.co/blog/analysis/fbi-confirms-north-korea-s-lazarus-group-as-hackers-behind-100-million-harmony-horizon-bridge-theft}
\BIBentrySTDinterwordspacing

\bibitem{poiFail}
\BIBentryALTinterwordspacing
AnChain.AI, ``Why {D}id {R}ailgun's {P}roof of {I}nnocence {F}ail? {A} {D}eep {D}ive into {P}rivacy {P}rotocol {V}ulnerabilities,'' 2025. [Online]. Available: \url{https://www.anchain.ai/blog/railgun-proof-of-innocence}
\BIBentrySTDinterwordspacing

\bibitem{stablecoins}
\BIBentryALTinterwordspacing
Chainway, ``Stablecoins 101: {B}ehind crypto’s most popular asset,'' 2022, accessed: 2025-03-11. [Online]. Available: \url{https://www.chainalysis.com/blog/stablecoins-most-popular-asset/}
\BIBentrySTDinterwordspacing

\bibitem{poseidon}
\BIBentryALTinterwordspacing
L.~Grassi, D.~Khovratovich, C.~Rechberger, A.~Roy, and M.~Schofnegger, ``Poseidon: A new hash function for zero-knowledge proof systems,'' in \emph{Proeedings of the 30th {USENIX} Security Symposium}, 2021. [Online]. Available: \url{https://www.usenix.org/system/files/sec21-grassi.pdf}
\BIBentrySTDinterwordspacing

\bibitem{trusted_setup}
\BIBentryALTinterwordspacing
V.~Nikolaenko, S.~Ragsdale, J.~Bonneau, and D.~Boneh, ``Powers-of-tau to the people: Decentralizing setup ceremonies,'' in \emph{Applied Cryptography and Network Security}, 2024. [Online]. Available: \url{https://link.springer.com/chapter/10.1007/978-3-031-54776-8_5}
\BIBentrySTDinterwordspacing

\bibitem{tornadoBlog}
\BIBentryALTinterwordspacing
``Introducing {P}rivate {T}ransactions {O}n {E}thereum {NOW}!'' 2019. [Online]. Available: \url{https://tornado-cash.medium.com/introducing-private-transactions-on-ethereum-now-42ee915babe0}
\BIBentrySTDinterwordspacing

\end{thebibliography}
    
\end{document}